\documentclass{article}

\usepackage{arxiv}

\usepackage[utf8]{inputenc} 
\usepackage[T1]{fontenc}    
\usepackage{hyperref}       
\usepackage{url}            
\usepackage{booktabs}       
\usepackage{amsfonts}       
\usepackage{nicefrac}       
\usepackage{microtype}      
\usepackage{lipsum}
\usepackage{subfig}
\usepackage{natbib}
\usepackage{graphicx}
\usepackage{booktabs}
\usepackage{setspace}
\usepackage{pdfpages}
\usepackage{todonotes}
\usepackage{booktabs}
\usepackage{multirow}
\usepackage[normalem]{ulem}
\useunder{\uline}{\ul}{}

\usepackage[page]{appendix}

\title{Images and Misinformation in Political Groups: Evidence from WhatsApp in India}
\author{
 Kiran Garimella  \\
  MIT\\
  \texttt{garimell@mit.edu} \\
  \And
 Dean Eckles \\
  MIT\\
  \texttt{eckles@mit.edu} \\
}

\begin{document}
\date{}
\maketitle
\begin{abstract}

WhatsApp is a key medium for the spread of news and rumors, often shared as images. We study a large collection of politically-oriented WhatsApp groups in India, focusing on the period leading up to the 2019 Indian national elections. By labeling samples of random and popular images, we find that around 13\% of shared images are known misinformation and most fall into three types of images. Machine learning methods can be used to predict whether a viral image is misinformation, but are brittle to shifts in content over time. 

\end{abstract}


\section{Key Takeways}

\begin{itemize}
    \item We collected a large dataset of images from thousands of public WhatsApp groups in India. We annotated a sample of these images with the help of journalists.
\item Based on this annotation, we find that image misinformation is highly prevalent on WhatsApp public groups, making up up to 13\% of all images shared on these groups.
\item We quantify how images are being used to spread misinformation by categorizing the types of image misinformation, finding three main classes: images taken out of context, photoshopped images, and memes.
\item Based on our findings, we developed machine learning models to detect misinformation. While the results can sometimes appear promising, these models are not robust to changes over time.
\item This is the first work to study misinformation on WhatsApp and in the form of images at scale. Given the increasing importance of WhatsApp and other encrypted messaging services in our lives, such findings help in developing technology and policy.

\end{itemize}

\section{Argument \& Implications}
While group messaging services — most notably WhatsApp — clearly provide value to billions of people, misinformation spreading there has been the proximal cause of social unrest and violence. However, it has been difficult to systematically study misinformation on WhatsApp because of the semi-closed nature of the platform; thus, there is a substantial gap in our knowledge of the prevalence, typology, and detectability of misinformation on WhatsApp. Also, given that a large fraction of the messages shared on WhatsApp are images, understanding the prevalence of misinformation in such modalities is an important task. This paper tackles several novel problems, each of which have their own challenges.

\paragraph{Challenges in obtaining data.}
Collecting and characterizing data at the scale we did is not trivial. Our data consist of publicly shared WhatsApp messages, specific to India, and hence are subject to various biases. We may be missing, e.g., fringe, extreme content, which might not be discussed in open groups. There are likewise important ethical considerations in obtaining such data, for which there are no agreed upon norms in the research community. Nonetheless, this new data allows us to  better study the prevalence of image misinformation, and macro patterns of coordination across groups.

\paragraph{Challenges in annotation.}
We worked with journalists to annotate a sample of images, which included over 800 misinformation images. We estimate that  approximately 13\% of image shares in our dataset are misinformation. The annotation of image misinformation is challenging in large part because it needs expertise in fact checking and an image can be misinformation or not based on the context in which it was shared and the time when it was shared.
We built special interfaces to enable the annotation that include context while protecting the privacy of the users. However, our annotation only covers a very small subset (0.1\%) of all the images in our dataset, though we oversampled highly shared images. Even with such a small set, the annotations help in the characterization of the types of misinformation images, which in turn helps us understand the underlying motivations and potential methods to prevent the spread of such misinformation.

\paragraph{Challenges in detecting image misinformation.}
We characterize the prominent types of image misinformation and find that three categories of images make up almost 70\% of misinformation shared in our dataset. Of those, detecting  out-of-context images might be feasible using existing state-of-the-art techniques. We can easily recover images which are near duplicates, and can identify if an image has been shared in the past and flag images for potential misinformation if shared out of context. Identifying photoshopped images is hard, even manually (\cite{nightingale2017can}). State-of-the-art image processing techniques exist to detect manipulated JPEG images (\cite{wu2019mantra}). We tried using these techniques to identify if an image has been digitally manipulated, but these tools were not particularly useful here. These tools confuse genuine photoshopped images with memes, which are also ``doctored" in a sense, with someone adding a piece of text or a color when creating a meme.

We also found that misinformation images have certain peculiar characteristics, in terms of their sharing and spreading characteristics compared to non-misinformation images (cf. \cite{vosoughi2018spread}, \cite{friggeri2014rumor}). We trained machine learning models to identify misinformation and show that though most models do better than a random guess, they are not close to being deployable. Notably, we find that content-based classifiers are not robust to the choice of training data, with performance degrading across time periods.

\paragraph{What next?}
How do the insights presented in this paper make a difference? Firstly, the typology of image misinformation and the insights presented in our paper will be helpful to regulators, technology companies, civic campaigners, and those interested in media literacy and political communication. There are three main stakeholders in solving the issue of misinformation.
\begin{enumerate}
    \item Technology: Since we find that  out-of-context images do constitute a large fraction of image-based misinformation, a simple feasible technical solution to tackle part of the problem is to create a common repository of old images and fact-checked images, which could be queried to check if an image was already shared in the past.
\item Users: One important missing piece in our work is to understand why users share misinformation. More qualitative (e.g., \cite{banaji2019whatsapp}) and psychological (e.g., \cite{pennycook2019lazy}) studies are needed. These studies could be used to educate users and to develop interventions to solve the problem.

\item Platforms: Ultimately, since WhatsApp is a closed platform, any technique that is developed to detect and debunk misinformation can only be implemented by WhatsApp/Facebook, which should build tools necessary to help users identify misinformation effectively.
After all, there may not be a concrete solution just based on one of these three stakeholders that solves the misinformation problem. A more plausible solution could be a mix of technical, social and policy based ideas.

\end{enumerate}

\section{Findings}

\textbf{Finding 1:} Image misinformation is highly prevalent in public political WhatsApp groups.

In order to understand the problem of image misinformation, we had 2,500 images manually annotated by three independent journalists. The 2,500 images consists of two sets: (i) a random sample of 500 images taken from our 1.6 million images (\textit{random}), and (ii) a set of 2,000 most shared images from our data (\textit{popular}). We sampled these two sets differently in order to get a sense of the underlying distribution of misinformation in the rest of the dataset. We asked the annotators to select from 4 options:  1. Misinformation, 2. Not misinformation, 3. Misinformation already fact checked, 4. Needs concrete fact checking (unclear) (see Methods for details). We also grouped together similar images into clusters, using an image hashing technique that identifies perceptually similar images (\cite{zauner2010implementation}).

First, we compute the prevalence of the various types of misinformation in both the samples (\textit{popular} and \textit{random}) and populations of interest. The first two columns of Table 1 show simple proportions of the two types of labeled samples given each majority label. As might be expected, previously fact-checked misinformation is much more common among the most popular images (that is, image clusters), compared with the random sample. In order to estimate the prevalence of these types of images in the broader set of images in the monitored groups, we used standard inverse probability weighting (IPW) estimators and associated standard errors that are widely used in survey sampling to estimate these proportions (\cite{sarndal2003model}, \cite{lumley2004analysis}). First, we can estimate prevalence among all unique images (i.e., image clusters). Because this gives equal weight to all images, the results are very similar to the random sample. We find that approximately 10\% of image shares are obviously (to journalists) misinformation, with 3\% more being misinformation that has been previously fact-checked. Clearly, misinformation makes up a substantial portion (13\%) of all image sharing in these political groups, even if we are unable to conclusively label many images as misinformation or not.

\begin{table}[ht]
\caption{Prevalence (\%) of image types.
The first two columns are for the two samples of labeled images. The second two use sampling weights to estimate prevalence in the population of all unique image clusters and the population of image shares.
The latter uses post-stratification on percentiles of number of times shared. Standard errors are given in parentheses.} 
\centering
\begin{tabular}{lllll}
  \hline
 & \textit{popular} & \textit{random} & clusters & shares \\ 
  \hline
factchecked & 7.0 (0.6) & 2.6 (0.7) & 2.6 (0.9) & 3.2 (0.7) \\ 
  misinfo & 18.8 (0.9) & 10.7 (1.4) & 9.3 (1.5) & 11.1 (1.2) \\ 
  not misinfo & 39.4 (1.2) & 43.5 (2.2) & 46.2 (2.6) & 43.5 (2.1) \\ 
  unclear & 34.9 (1.1) & 43.3 (2.2) & 41.8 (2.6) & 42.2 (2.1) \\ 
   \hline
\end{tabular}
\label{tab:prevalence}
\end{table}

\begin{figure}[ht!]
\centering
\includegraphics[width=\textwidth, ]{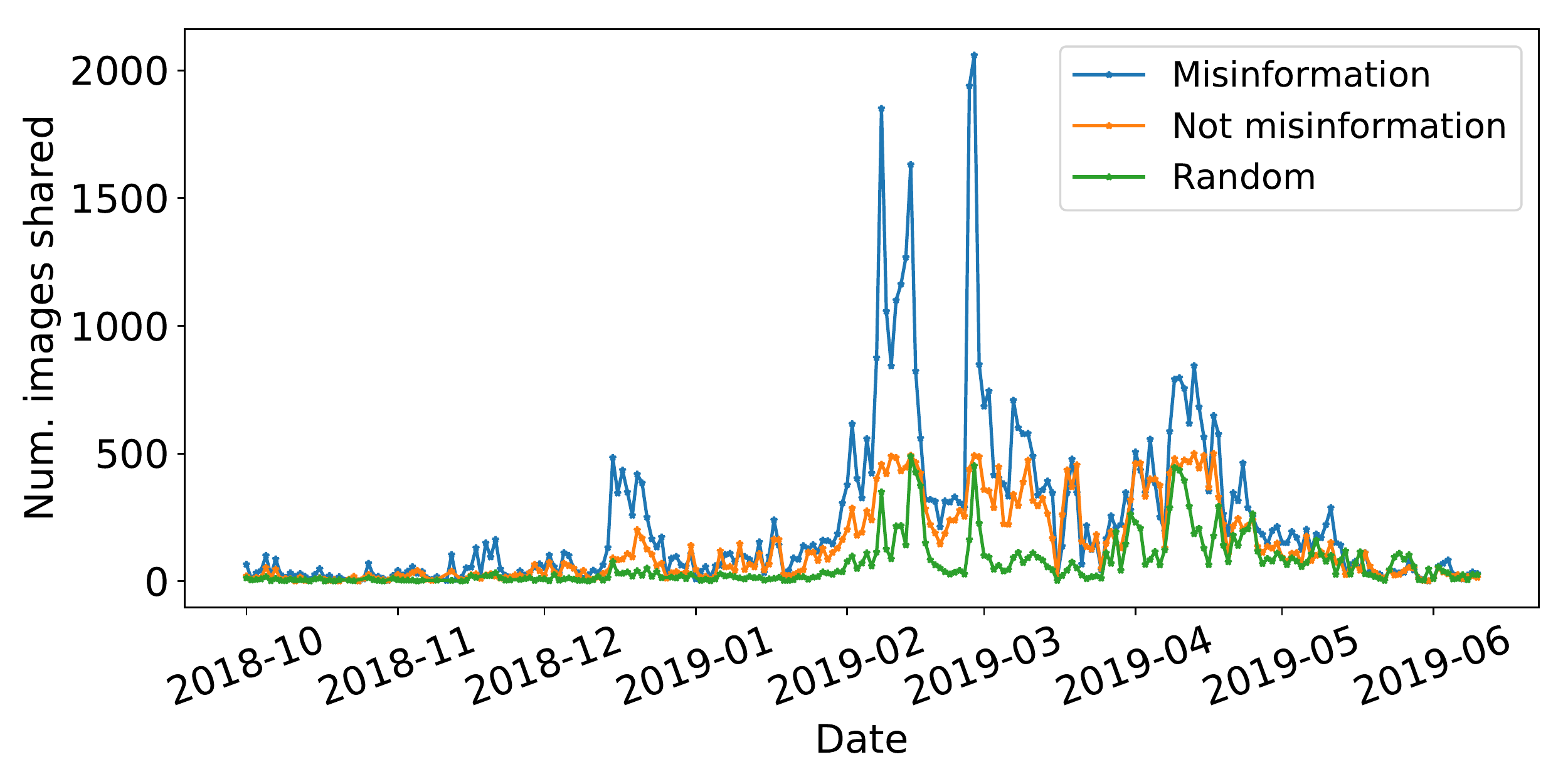}
\caption{Count of the number of images from the different sets shared over time. We see clear peaks in the shares of misinformation during the India/Pakistan tensions in February/March, and elections in April.}
\label{fig:over_time}
\end{figure}

Figure~\ref{fig:over_time} shows the trends the shares of misinformation. We see clear peaks in the values of the number of misinformation images shared: a. Sabarimala protests (late December), b. Terrorist strike in Kashmir, India (mid February), c. India/Pakistan tensions (late February, early March), and d. The elections (April). At their peak during the India/Pakistan conflict, over 2000 instances of misinformation images were shared on a day. Even during the election campaigns, roughly 800 images were shared per day. Note that we only annotated a very small fraction of misinformation from the complete set. So, this is still a small fraction of all misinformation in our data, and an even smaller fraction of misinformation on WhatsApp. But these numbers give us a sense of how prevalent the problem of misinformation is.

\textbf{Finding 2}: Most image misinformation falls into three categories: old images taken out of context, memes and photoshopped images.

Qualitative analysis of the image types reveals that the misinformation covers a wide spectrum of images including historic \& religious myths, nationalism and political memes. Many images convey a sense of immediacy, and shock value (violent/graphic images) which makes them spread faster. We created a typology of the various types of image based misinformation. 
This will help us understand the different types of deception used. 
Based on manual coding, we were able to identify 3 main types of images: 
(i) Old images that are taken out of context and reshared,
(ii) images which are manipulated/edited, 
(iii) memes - funny, yet misleading images, containing incorrect stats or quotes, and,
(iv) other types of fake images, consisting of health scares, fake alerts, superstitions, etc. 
We should note that this is not an exhaustive list of categories and could be specific to our dataset.

\begin{figure*}[ht!]
\centering
\begin{minipage}{.32\linewidth}
\centering
	\subfloat[]{\label{fig:photoshopa}\includegraphics[width=\textwidth, height=\textwidth, clip=true, trim=0 0 0 0]{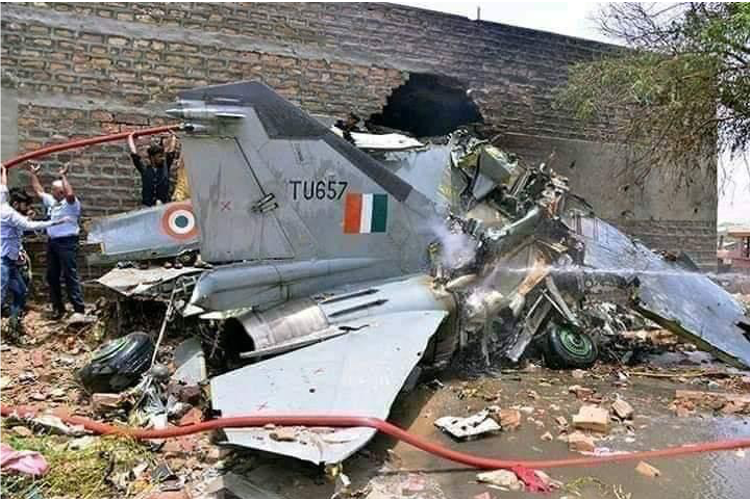}}
\end{minipage}%
\begin{minipage}{.32\linewidth}
\centering
	\subfloat[]{\label{fig:photoshopa}\includegraphics[width=\textwidth, height=\textwidth, clip=true, trim=0 0 0 0]{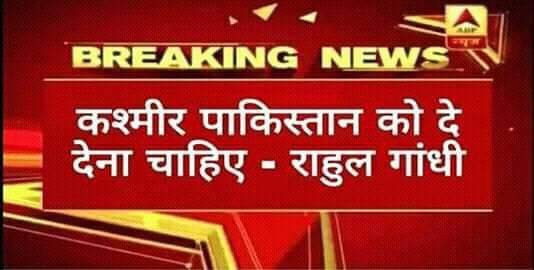}}
\end{minipage}%
\begin{minipage}{.32\linewidth}
\centering
	\subfloat[]{\label{fig:photoshopa}\includegraphics[width=\textwidth, height=\textwidth, clip=true, trim=0 0 0 0]{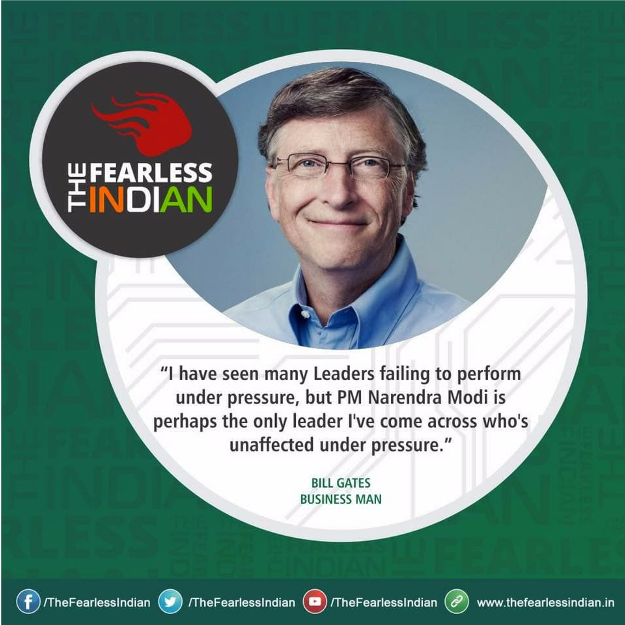}}
\end{minipage}%
\par\medskip
\begin{minipage}{.32\linewidth}
\centering
	\subfloat[]{\label{fig:photoshopb}\includegraphics[width=\textwidth, height=\textwidth, clip=true, trim=0 0 0 0]{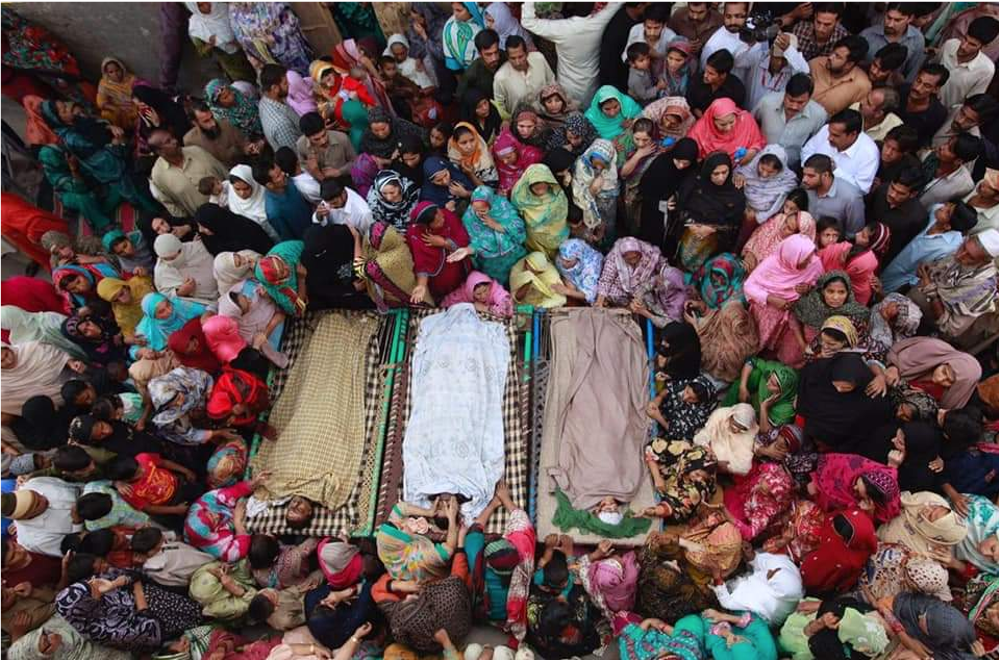}}
\end{minipage}%
\begin{minipage}{.32\linewidth}
\centering
	\subfloat[]{\label{fig:photoshopb}\includegraphics[width=\textwidth, height=\textwidth, clip=true, trim=0 0 0 0]{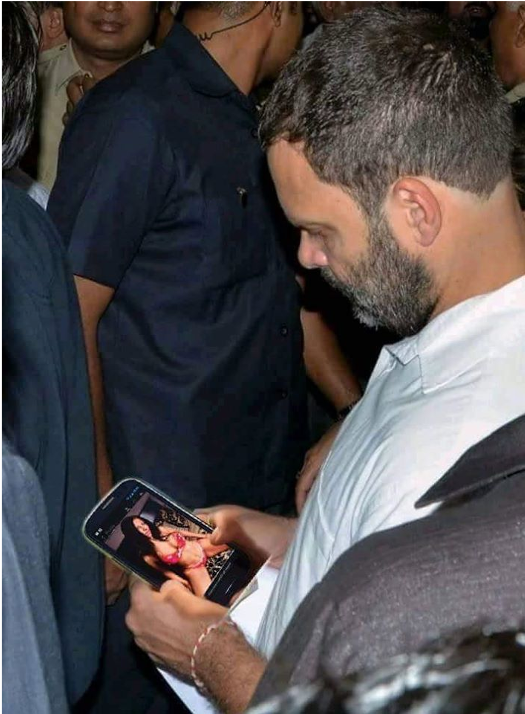}}
\end{minipage}%
\begin{minipage}{.32\linewidth}
\centering
	\subfloat[]{\label{fig:photoshopb}\includegraphics[width=\textwidth, height=\textwidth, clip=true, trim=0 0 0 0]{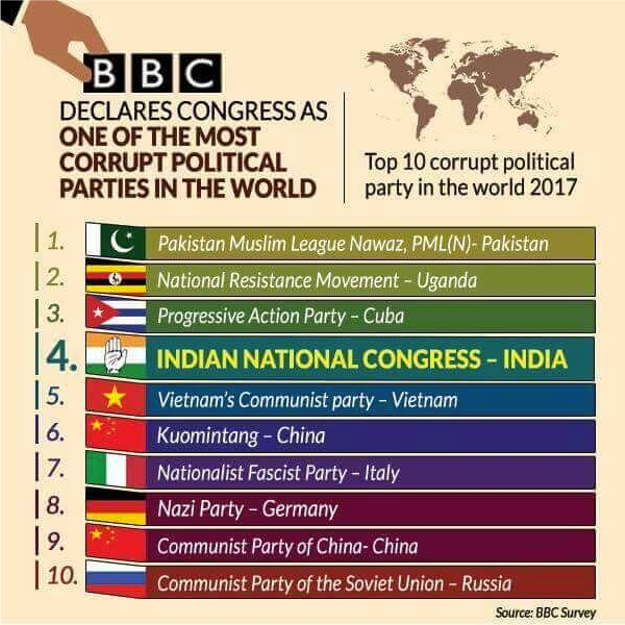}}
\end{minipage}%
	\caption{Types of misinformation. (a,d) Images taken out of context -- both these are old images claimed to be a result of India's attack on Pakistan; (b,e) Photoshopped images, (b)---doctored screen shot of a TV news program, (e)---of a politician browsing porn; (c,f) Fake quote by Bill gates, and a fake poll by BBC.}
	\label{fig:misinformation_types}
\end{figure*}

We find that old images that are taken out of context, also termed cheap or shallow fakes (Paris \& Donovan 2019) make up roughly 34\% of our misinformation image dataset. The next popular category is memes with fake quotes or statistics which make up 30\% of the images, followed by simple doctored (i.e., photoshopped) images, which make up roughly 10\% of the images. Note that just the top three categories of misinformation constitute over 70\% of the images. Examples of each category of such images is shown in Figure~\ref{fig:misinformation_types}. This characterization is similar to the qualitative characterization by (\cite{hemsley2018dimensions}).

The characterization of the images presented above helps us in two ways:
(i) To understand the motives behind posting. For instance, note that out of these three major types, two of them — photoshopped images and creating memes — require concerted effort to do at scale.
(ii) To develop automated techniques to identify misinformation in images. For instance, the out-of-context images are easy to detect using image hashing techniques, if they make use of images that appear in a database of content already shared in the past. Similarly, advances in computer vision techniques (\cite{wu2019mantra}) and image processing (\cite{lee2016recursive}) to identify text in images can be used to detect manipulated images. Since each of these types make up a significant portion of the misinformation, even trying to identify a subset of such images automatically would be immensely helpful.

\textbf{Finding 3}: Automatically identifying misinformation images is non-trivial and brittle.

Next, we try to answer the question: Can we identify whether an image is misinformation or not, based on its contents, how it spreads, where it spreads, and who spreads it? To that end, we built machine learning models based on various features extracted from our data. We also check whether some of the sub-categories of image misinformation presented above are easier to detect automatically.

We consider a binary classification task of predicting if an image is misinformation or not. The class distribution is balanced using the Synthetic Minority Oversampling Technique (\cite{chawla2002smote}). So a random prediction would have an accuracy of 0.5. For this task, we create a wide range of features (i.e., predictors), which primarily come under two categories: (i) image metadata features: how it spreads, where and when it spreads, and who spreads it, etc., and (ii) image content features: properties inherent to the content of the image. See the Methods section for details on the features. We compare between three sets: \textit{misinfo} (misinformation images labeled by experts), \textit{not misinfo} (images labeled specifically as not misinformation by experts), and \textit{random} (randomly sampled unlabeled images).

Table~\ref{tab:metadata_features} shows the accuracy of a classifier making use of just metadata features.
We make a few observations:
Detecting misinformation images from \textit{random} images is easier, with a much higher accuracy;
Even for \textit{not misinfo}, most features perform slightly better than random, with the web entities feature outperforming the others;
Combination of all the features does not improve the performance much.
These numbers indicate that there is a weak signal in the way misinformation images spread, being posted in a certain set of groups by a certain set of users.

\begin{table}[]
\centering
\caption{Classification using metadata features}
\begin{tabular}{c|c|c}
\hline
           & \textit{misinfo}\ vs. \textit{not misinfo} & vs. \textit{random} \\
\hline
Groups  & 0.65      & 0.82    \\  
Users  & 0.65      & 0.78    \\  
Time  & 0.57      & 0.77    \\  
Web domains & 0.59          & 0.75        \\
Web entities & 0.72            & 0.84       \\
Overall & 0.67            & 0.85       \\
\hline
\end{tabular}
\label{tab:metadata_features}
\end{table}

Next, we compare the performance of our models built using features extracted from the content of the image. These features were extracted using computer vision techniques and other publicly available tools like the Google Cloud Vision API to extract objects and faces from the image. The best accuracy we could obtain from this task was just slightly better than random (accuracy 0.57, 0.71 for \textit{misinfo} vs. \textit{not misinfo} and vs. \textit{random} respectively, see results in Table~\ref{tab:google_vision}). On the other hand, a pre-trained ResNext model performs very well, even reaching over 0.95 in some cases.

\begin{table}[]
\centering
\caption{Classification using: Top: Google Vision API, Bottom: pretrained ResNext model.}
\begin{tabular}{c|c|c}
\hline
             & vs. \textit{not misinfo} & vs. \textit{random} \\
\hline
Safe search  & 0.57            & 0.58       \\
Objects      & 0.55            & 0.63       \\
Text         & 0.57            & 0.71       \\
Face         & 0.57            & 0.53	    \\
Color       & 0.52            & 0.55	    \\
\hline
\hline
ResNext	(without text) &   0.78 & 0.92 \\
ResNext	(with text) & 0.78 & 0.96 \\
\hline
All combined  &	0.78 & 0.95 \\
\end{tabular}
\label{tab:google_vision}
\end{table}

To ensure that the results we obtained are robust, we split the images into different sets based on when they were shared. We find that when testing on images with different time periods, metadata features are robust and do not change much irrespective of what the training/test dataset is. However, on the other hand, image content related features are fragile and have quite a bit of variance, going as low as 0.55 from 0.95. A lot of work in the area of predicting "fake news" regularly reports over 85\% accuracy (\cite{gupta2013faking}, \cite{monti2019fake}, \cite{khattar2019mvae}). Our results indicate the importance of testing the robustness of such approaches, especially for blackbox models.

Finally, we checked whether our machine learning models are better at identifying the three major types of image misinformation: out-of-context images, memes, and manipulated images. We find that identifying out-of-context images works better by using the web domains feature with an accuracy of 0.78 for the \textit{misinfo} vs. \textit{not misinfo} task (compared to 0.59 for all the images). This is expected since this feature captures where else on the web this image appeared in the past, and  out-of-context images are typically more likely to have appeared in low quality or fact checking websites, which makes them easily discernible. We do not observe any significant differences in accuracy for detecting manipulated images or memes, even using features specific to those images, whether features from a recent paper that detects potentially manipulated pixels (\cite{wu2019mantra}) or features about text in the image (see Methods).

\textbf{Finding 4}: Political parties differ in the levels of misinformation shared in the groups.
We find that in our dataset, BJP has the highest fraction of misinformation images (8\%), where as Congress is slightly lower (3\%).

\begin{figure}[ht]
\centering 
\includegraphics[width=0.55\textwidth]{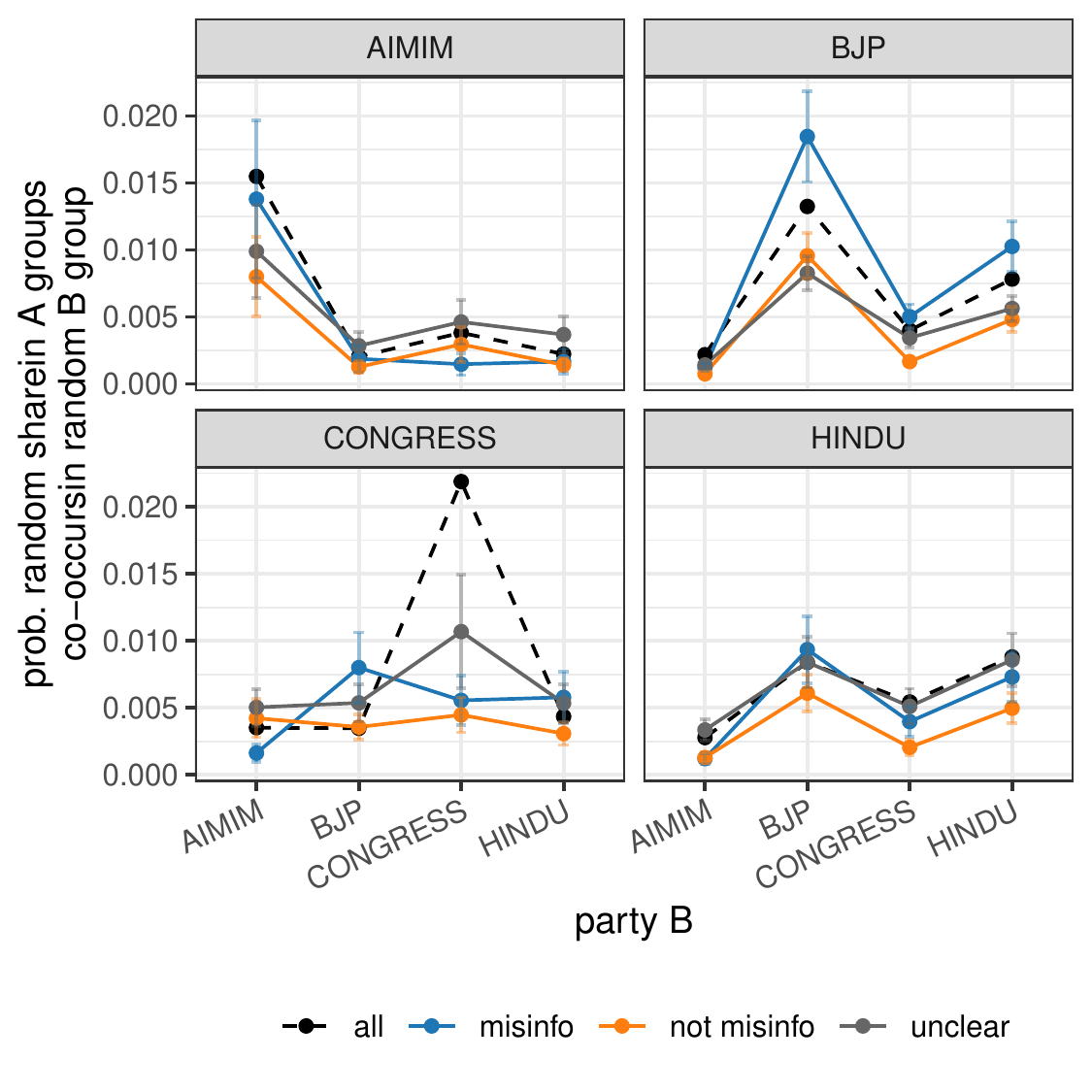}
	\caption{Cosharing of images in groups associated with the same and different political party. For labeled images, these are IPW estimates using the known sampling probabilities with error bars showing standard errors reflecting this.}
	\label{fig:party_cosharing_labels}
\end{figure}

We computed the intersection of the top 100 most shared images within each party to check if the major parties share similar misinformation or if they have their own set of misinformation images.
We were mainly interested in 4 groups: The two big national parties, BJP \& Congress, and the two biggest religions, Hindu and Muslim groups.
BJP/Congress have a 31\% overlap, and Hindu/Muslim have a 21\% overlap, indicating that these groups share little in common.
We also computed the same overlap for general images (not only misinformation), and these numbers are even lower at 10\% and 6\% respectively.
There seems to be some common misinformation shared by these two groups. To examine the co-occurrence of images beyond the most shared images, we consider the probability that a random share from party A is an image that also appears in a random group of party B. Figure \ref{fig:party_cosharing_labels} shows these cosharing rates both for all images and for images of each label type;  the estimates for the labeled data are inverse-probability weighted (IPW). Unsurprisingly, images are more likely to appear in other groups of the same party compared with other parties. This carries over to ideologically related groups, such as cosharing between BJP and Hindu groups, and this pattern appears, albeit with less statistical precision, when considering images identified as misinformation.

\section{Methods}

\subsection{Data collection}
Our analysis relies on a large dataset that we obtained from public groups on WhatsApp. Specifically, we focus on public groups discussing Indian politics. Lists of such public groups are advertised through dedicated websites and social media. The use of public groups has become a de facto way for political parties to reach a new audience, who are only available through WhatsApp. Surveys have shown that such groups are very popular in India (\cite{lokniti2018}) and Brazil (\cite{reuters2019report}). We obtained groups related to politics in India by manually searching for keywords related to all major national \& regional political parties and religious groups and their prominent political leaders, looking for data on Facebook, Twitter and Google. All the related chat.whatsapp.com links that were obtained were then manually screened and we removed any link that was not deemed to be politically relevant.

We joined and obtained data from over 5,000 political groups using the tools provided by (\cite{garimella2018whatsapp}). From these groups, we obtained all the text messages, images, video and audio shared in the groups. For each group, we also know the members of the group, and the admin(s) of the group.We collected over 5 million messages shared by over 250,000 users over a period of 9 months, from October 2018 to June 2019. Given high profile events such as the Indian national election, India/Pakistan tensions, etc, data from political groups during this period is extremely insightful. The data covers all the major national and regional parties, and affiliated religious groups, with over 10 languages represented. Roughly half of the content is in Hindi, followed by English, Telugu and other regional languages. Roughly 41\% of the data is text, 35\% images, and 17\% video. In this paper, we only focus on images, since images were a dominant modality of information sharing on WhatsApp. In total, our dataset consists of 1.6 million images.

An important point to note about our data is the biases in sampling the groups. Since WhatsApp does not provide a way to search available groups, it is difficult to estimate the bias in our sample. A large fraction of the groups in our data (24.4\%) apparently belong to the BJP (the party currently in power) and other unofficial religious groups losely supporting the BJP. We could not find many groups from the main opposition party, Indian National Congress, which only accounts for 5\% of the groups in our data. This bias could just reflect the lack of a coordinated WhatsApp strategy for the Congress.

\subsection{Annotating the images for misinformation}

\begin{figure}[ht!]
\centering
\begin{minipage}{.49\linewidth}
\centering
	\subfloat[]{\label{}\includegraphics[width=\textwidth, height=0.8\textwidth]{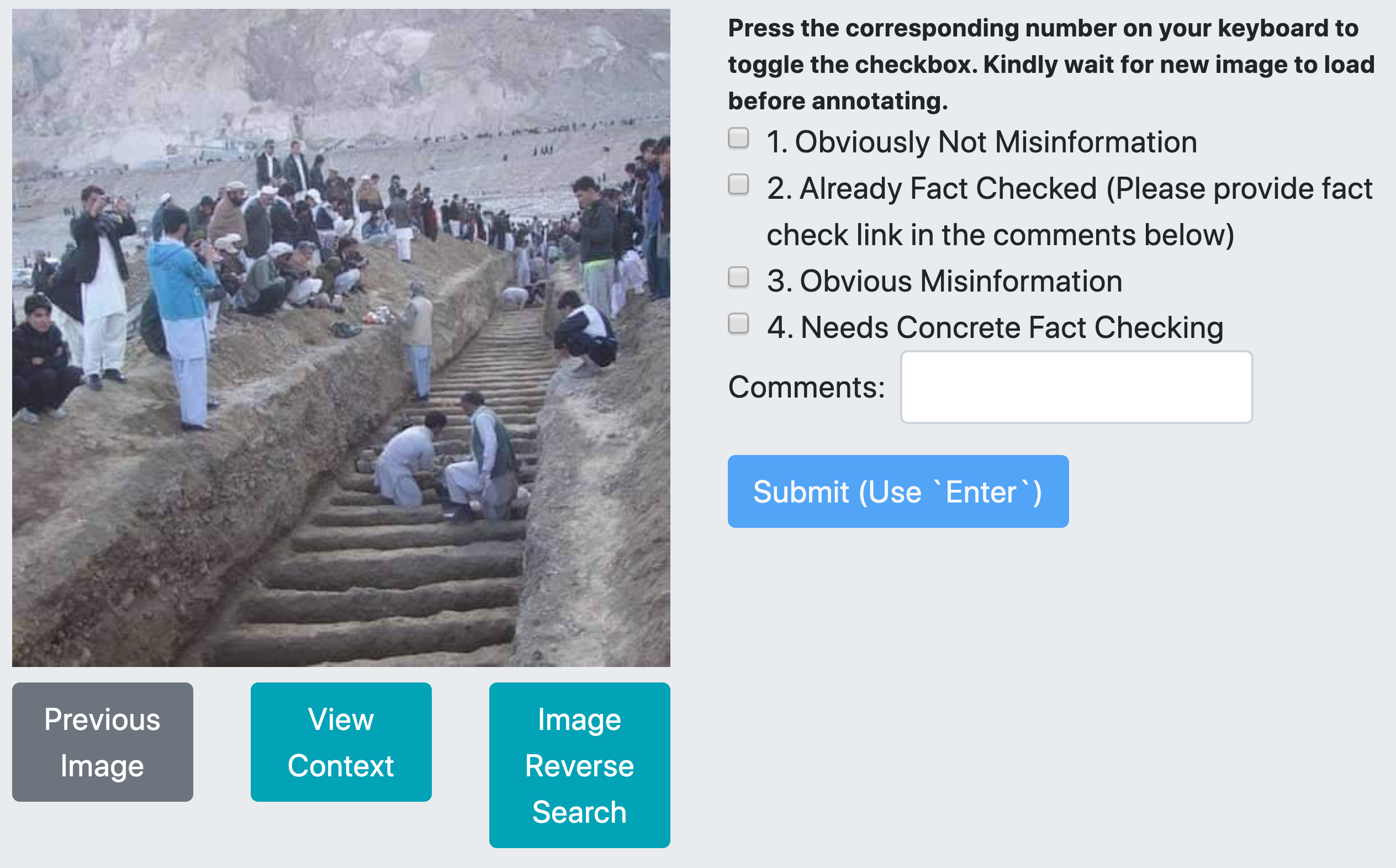}}
\end{minipage}%
\begin{minipage}{.49\linewidth}
\centering
	\subfloat[]{\label{}
	\includegraphics[width=\textwidth, height=0.8\textwidth]{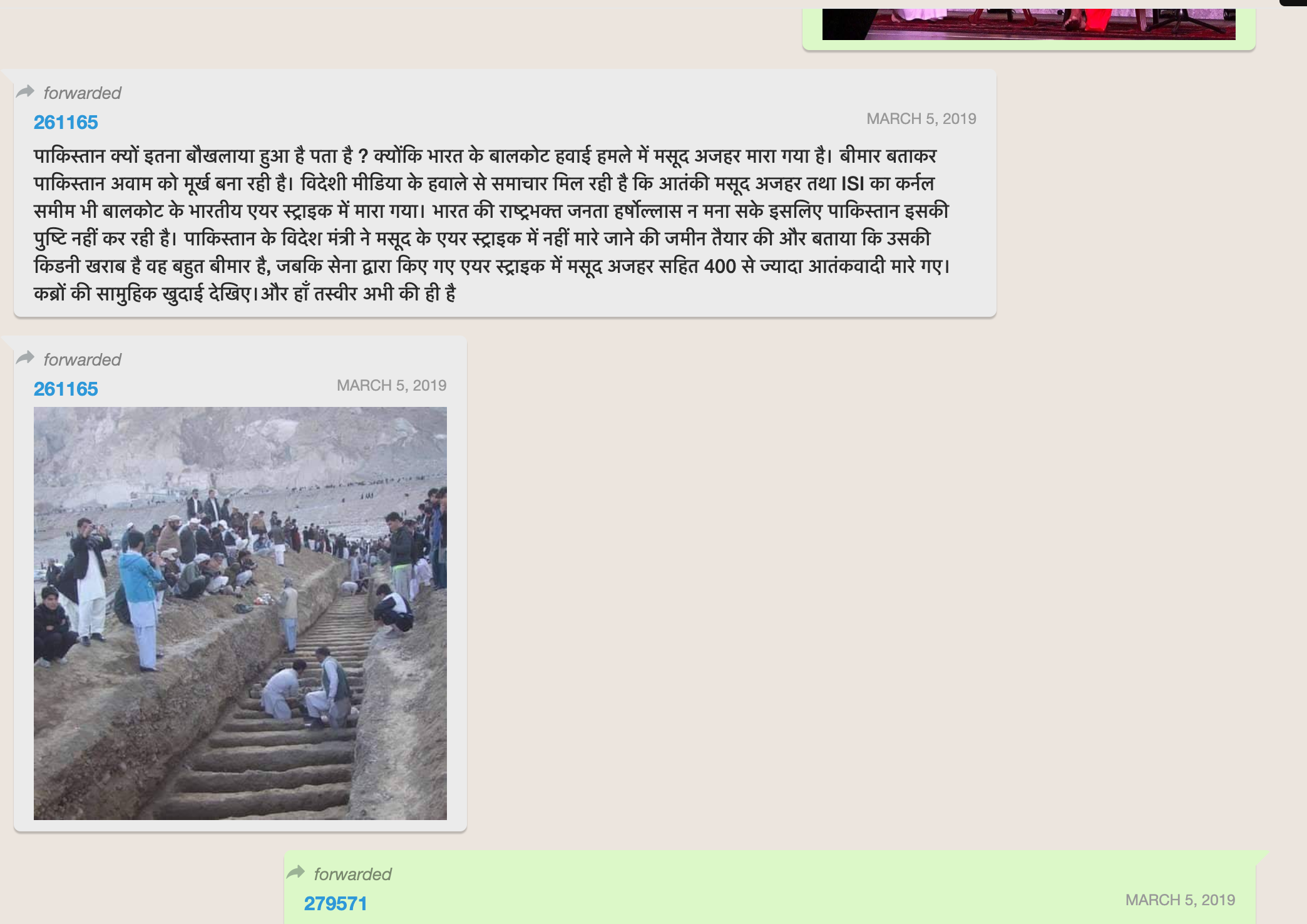}}
\end{minipage}%
\par\medskip
	\caption{(a) Annotation portal used for annotating misinformation images, (b) Image shown in the chat context in which it was shared.}
	\label{fig:annotation_portal}
\end{figure}

We manually annotated a sample of 2,500 images with the help of three professional journalists. The journalists are experts in fact checking and regularly report on digital platforms. The 2,500 images were chosen to contain both a random sample of 500 images, and a set of 2,000 popular (most shared) images from our data. This effectively oversamples popular images while also enabling estimating prevalence in the full set of images. We built an annotation portal specifically for the journalists to view and annotate the images. Figure~\ref{fig:annotation_portal} shows the annotation portal. The portal shows an image and the annotators select from a set of four available options. Tools to check the image in the chat context in which it was shared, and to reverse image search were provided.

We asked the annotators to select from four options: 1. Obviously not misinformation 2. Obvious misinformation 3. Already fact checked to be false 4. Needs concrete fact checking. The annotators could select only one option. In case an image is both misinformation and is fact checked, they selected the latter. The design of the options was due to the following considerations: a. We wanted to get a sense of the fraction of images that were already fact checked, b. Fact checking is an extremely time consuming operation for certain cases, and since we employed professional journalists, we could not ask them to spend a lot of time on each image,given that they had to annotate 2,500 images. So we gave them option 4 above, which they were asked to select only if they were highly uncertain.

The annotators had the following instructions: For each image, do you know if the image contains obvious misinformation (or not)? If it is not clear, first click on the "View Context" button, to show the image in the chat context along with the text with which it was shared (Figure~\ref{fig:annotation_portal} (b)). Next, check if it has already been fact checked by clicking on the "Image Reverse Search" option to check where it has appeared on the web (using Google reverse image search). If you find the image already fact checked to be false in popular fact checking sites, please report where it was fact checked. If the image has not appeared on the web and is hard to factcheck quickly, please mark the 4th option, needs concrete fact checking. Annotating images for misinformation is an extremely hard and time consuming task, even for professionals. Images do not necessarily indicate misinformation with our context. For example, the image in Figure~\ref{fig:annotation_portal} (a) might only be misinformation in the context claiming that it shows the burial of dead bodies of terrorists killed in an Indian airstrike. Sometimes a group of images are shared together,and only one of them might be misinformation. The other may not be false in isolation but when seen together they contribute to misinformation. So a fake image in one context could be completely benign in another context. There are also cases where images could be misinformation or not depending on when they were shared, with the context implied due to the time period. Thus, we had to go through a rigorous process of using experts, and further cleaning the results of the annotation. 

For getting a dataset of misinformation images, we combine the obvious misinformation and fact checked to false categories. The inter-annotator agreement (Fleiss' Kappa) for the three annotators and three classes (misinformation/not misinformation and needs concrete fact checking) is 0.42 indicating a moderate level of agreement. In most cases, the disagreement comes from labelling (not) misinformation as needs concrete fact checking. The Fleiss Kappa for just the two classes (misinformation/not) is 0.78. In order to make sure that our final labeled dataset is of high quality, we did not consider the small set of cases with obvious disagreement (one journalist labeling it as misinformation while the other not). This gave us 605 images which were annotated to be false.

\subsection{Fact checked images}
We also collected images which were fact checked on popular fact checking websites from India. By scraping all the images present on these websites, we obtained close to 20,000 images. By matching the images from fact checking websites to the millions of images from our WhatsApp dataset using image matching techniques, we can find examples of fact checked images in our data. To do this, we used state of the art image matching techniques developed and being used at Facebook. The matching works by computing a 256-bit hash for each image. The hashing is done using a technique called perceptual hashing, which can detect near similar images, even if cropped differently or having small amounts of overlaid text overlaid. This gave us roughly 200 images which were fact checked to be false. We combined this with the above set manually labelled by us to obtain our final set of misinformation images.

\subsection{Image clustering}
\begin{figure}[ht]
\centering
\begin{minipage}{.24\linewidth}
\centering
	\subfloat[]{\label{}\includegraphics[width=\textwidth, height=\textwidth]{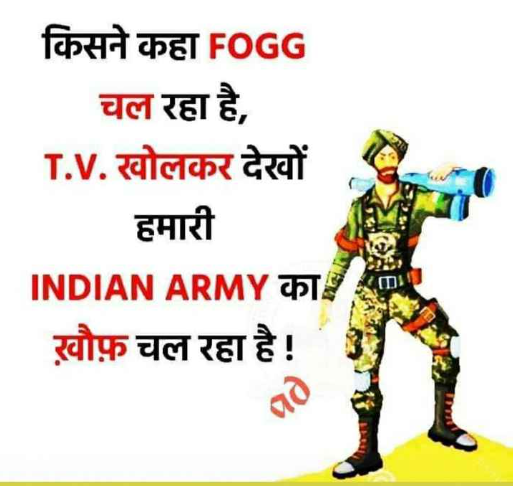}}
\end{minipage}%
\begin{minipage}{.24\linewidth}
\centering
	\subfloat[]{\label{}\includegraphics[width=\textwidth, height=\textwidth]{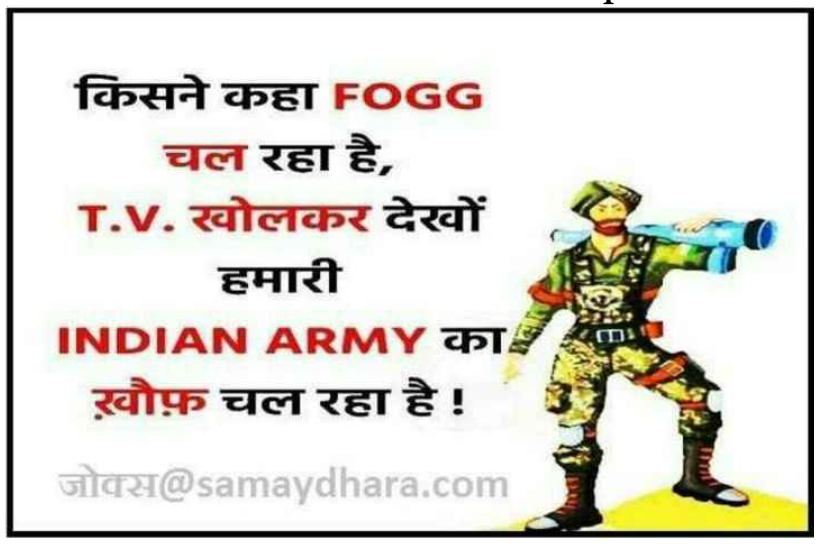}}
\end{minipage}%
\begin{minipage}{.24\linewidth}
\centering
	\subfloat[]{\label{}\includegraphics[width=\textwidth, height=\textwidth]{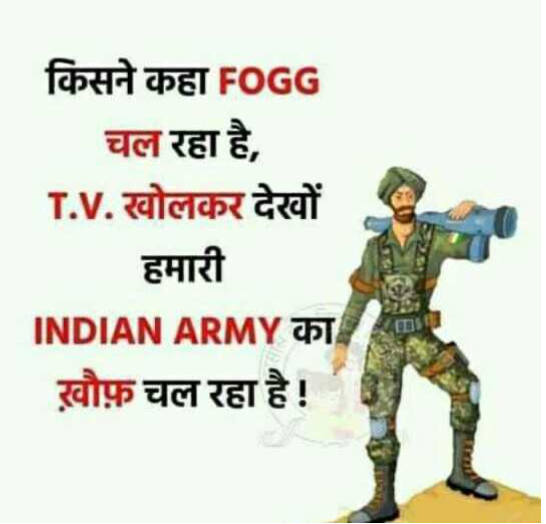}}
\end{minipage}%
\begin{minipage}{.24\linewidth}
\centering
	\subfloat[]{\label{}\includegraphics[width=\textwidth, height=\textwidth]{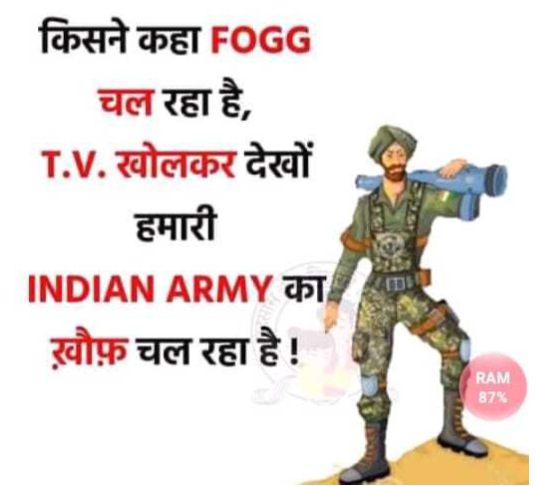}}
\end{minipage}%
\par\medskip
	\caption{Four examples of the same image with different characteristics, clustered together using Facebook PDQ hashing. a. image with some text, b. image with a border added, c. image with a different background, and, d. image with a different color filter applied and a red logo added.}
\label{fig:image_clustering}
\end{figure}

Though images are hard to manipulate, we observe many instances of similar images with slight changes (e.g. see Figure~\ref{fig:image_clustering}) in our dataset. In order combine variants of a similar image, we used the same image hashing algorithm (Facebook PDQ) to obtain hashes, and cluster the hashes together using the DBSCAN clustering algorithm (\cite{ester1996density}). This step ensures that we group together images which have similar content, and hence consider all variants of the image as a single cluster. A similar technique was used by (\cite{zannettou2018origins}) to measure the spread of memes in the US. Next, for a given annotated misinformation image,we include all the variants of images which were clustered together with the image as potential misinformation. This gave us around 40 thousand image instances corresponding to the 800 clusters, shared by over 13 thousand users.

\subsection{Prevalence estimation}

Note that we only annotated 2,500 images (2000 popular, and 500 random) from our entire dataset of millions of images. In order to estimate the prevalence of these types of images in the broader set of images in the monitored groups, we combine these samples for estimation and inference. In particular, the combined sample has positive (but heterogeneous) probability of including any image, so we can use standard inverse probability weighting (IPW) estimators and associated standard errors that are widely used in survey sampling to estimate these proportions (\cite{sarndal2003model},
\cite{lumley2004analysis}). More interestingly, we can also estimate prevalence among image shares, which reflects the chance that a random image share is of each type. For these estimates, we additionally post-stratify on percentiles of number of image shares to increase precision.

\subsection{Machine learning models}

To detect whether an image contains misinformation or not, we built machine learning models based on various features extracted from our data. The features can be grouped into two types: (i) Metadata features, which look at only information about the spread of the image in our data, and (ii) content features, which look at the content of the images.

Metadata features: We used various types of metadata features of the images such as:
(I) Spread features: (i) Groups in which the images were shared, (ii) users who shared the images, and (iii) how many times the image was shared.
(II) Time features: (i) total time span of the image (difference in days between the first and last share), and (ii) average time between shares (a measure of virality).
(III) Web features: (i) We used reverse Google image search to get a list of domains on which the image appears and used the list of domains as a feature, (ii) We also used the "web entities" end point from the Cloud Vision API, which returns entities found on the web related to the image.

Now we check if there are any differences in the content of the images between the three classes.
We primarily used Google's Cloud Vision API to process and extract image content features.
Cloud Vision API provides tools to identify objects and text in the image effectively, and works quite well even for non English text. We obtained the features related to objects, text and faces from the image. Specifically, we extracted the following features:
(I) Safe Search features: detects whether the image contains violence, racy, adult, etc.
(II) Objects features: detects what objects are contained in the image.
(III) Text features: Identifies text in the image (OCR) and if text exists, we create a tf-idf vector of the bag of words from this text.
(IV) Face features: For images with faces, we extract facial features, like whether the person is angry, surprised, smiling, etc.
(V) Color features: Identify the dominant colors in the image.

In addition to the content features from Cloud Vision, we also obtained the fraction of pixels in an image that have been manipulated (\cite{wu2019mantra}) and used a pre-trained convolutional neural network (CNN) trained with Facebook's ResNext architecture. We used a publicly available model provided by PyTorch pre-trained on ImageNet. The model is trained to predict one of the 1,000 ImageNet classes, given an image. For each image, we just consider the last layer of the model, a 1000-dimensional vector, indicating the probability of the image belonging to one of the 1,000 ImageNet classes as the features. We also included the text that was posted along with the image. To do this, we first obtained Doc2vec embeddings for text before and after the image. This gave us a 200 dimensional feature vector for text before and after each image. We tested multiple classifiers using the above features and chose a Random Forest classifier since it provides the best accuracy. All results reported are the mean of a 10 fold cross validation.

\begin{figure}[ht]
\centering
\begin{minipage}{.49\linewidth}
\centering
	\subfloat[]{\label{}\includegraphics[width=\textwidth, ]{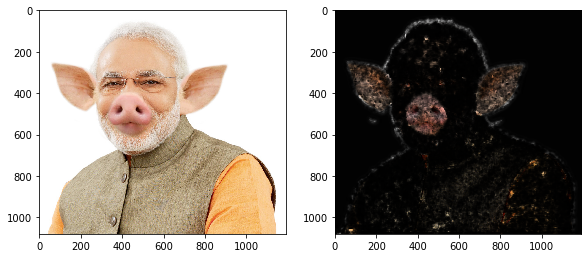}}
\end{minipage}%
\begin{minipage}{.49\linewidth}
\centering
	\subfloat[]{\label{}\includegraphics[width=\textwidth, ]{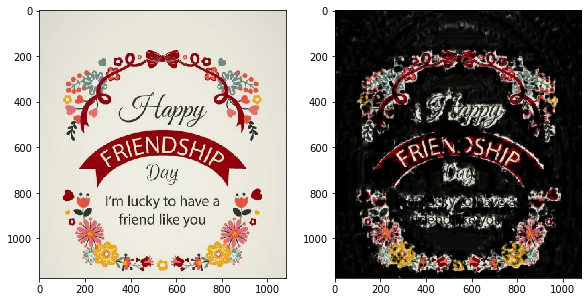}}
\end{minipage}%
\par\medskip
	\caption{Manipulated pixels detected using Matra-net (\cite{wu2019mantra}). The image on the left is clearly manipulated. The tool detects this manipulation and highlights the manipulated pixels. Using the fraction of manipulated pixels, we can identify this image is photoshopped automatically. However, the image on the right is a benign greeting, where some "manipulation" was done (adding text on a plain background). The tool identifies all the text and graphics added as manipulated pixels, and is unable to distinguish between these two types of manipulation.
}
\label{fig:image_manipulation}
\end{figure}

We also tested the classifiers on the three types of image misinformation: images taken out of context, memes and manipulated images. Our expectation was that these features, specifically engineered for a specific type of misinformation would help in detecting that subset of misinformation images. We hypothesize that the following category specific features would work well. (i)  Out-of-context images: Web domains. Web domains indicate the domains which are returned on a Google reverse image search for the image.  Out-of-context images are typically shared by low quality domains or by domains which might have fact checked the image. Typically such low quality or fact checking domains are not returned for other categories of images, (ii) Manipulated images: We used a state of the art computer vision technique, which detects pixels which might have been manipulated (\cite{wu2019mantra}). From this, we computed the fraction of pixels in an image which could have been manipulated. The intuition was that this fraction would be higher for manipulated images. However, the technique fails in our case since a lot of images which do not fall into this category are also falsely labeled as manipulated (see Figure~\ref{fig:image_manipulation} for an example), and, (iii) Memes: text on the image. Memes contain a lot of text on them. This category mostly contains memes with false quotes and statistics. The idea was that using text related features (e.g. tf-idf vectors), we could identify such false text.

\section{Appendix}

\textbf{Funding} K.G. is supported by a Michael Hammer postdoctoral fellowship administered by the Institute for Data, Systems \& Society at MIT. Some of this research was supported by a Junior Faculty Research Award by the MIT Sloan School of Management to D.E. We used some computing time provided by Amazon as part of an Amazon Research Award to D.E.

\textbf{Competing interests}
D.E. was previously an employee of Facebook and until 2019 had a significant financial interest in Facebook. D.E. has a significant financial interest in Amazon and Google. Some of D.E.’s research is funded by Amazon and The Boston Globe.

\textbf{Ethics}
This research was approved by MIT's Committee on the Use of Humans as Experimental Subjects, which included waiver of informed consent. The WhatsApp profile used for the data collection clearly states that we are a research organization and will be collecting the data for research.

\textbf{Copyright}
This is an open access article distributed under the terms of the Creative Commons Attribution License, which permits unrestricted use, distribution, and reproduction in any medium, provided that the original author and source are properly credited.

\textbf{Data Availability}
All the annotated images and their anonymized sharing patterns are made available here: https://zenodo.org/record/3779157.

\bibliography{biblio.bib}

\begin{thebibliography}{20}
\providecommand{\natexlab}[1]{#1}
\providecommand{\url}[1]{\texttt{#1}}
\expandafter\ifx\csname urlstyle\endcsname\relax
  \providecommand{\doi}[1]{doi: #1}\else
  \providecommand{\doi}{doi: \begingroup \urlstyle{rm}\Url}\fi

\bibitem[Nightingale et~al.(2017)Nightingale, Wade, and
  Watson]{nightingale2017can}
Sophie~J Nightingale, Kimberley~A Wade, and Derrick~G Watson.
\newblock Can people identify original and manipulated photos of real-world
  scenes?
\newblock \emph{Cognitive research}, page~30, 2017.

\bibitem[Wu et~al.(2019)Wu, AbdAlmageed, and Natarajan]{wu2019mantra}
Yue Wu, W~AbdAlmageed, and P~Natarajan.
\newblock Manipulation tracing network for detection and localization of image
  forgeries with anomalous features.
\newblock In \emph{CVPR}, 2019.

\bibitem[Vosoughi et~al.(2018)Vosoughi, Roy, and Aral]{vosoughi2018spread}
Soroush Vosoughi, Deb Roy, and Sinan Aral.
\newblock The spread of true and false news online.
\newblock \emph{Science}, 359\penalty0 (6380):\penalty0 1146--1151, 2018.

\bibitem[Friggeri et~al.(2014)Friggeri, Adamic, Eckles, and
  Cheng]{friggeri2014rumor}
Adrien Friggeri, Lada Adamic, Dean Eckles, and Justin Cheng.
\newblock Rumor cascades.
\newblock In \emph{ICWSM}, 2014.

\bibitem[Banaji and Bhat(2019)]{banaji2019whatsapp}
Shakuntala Banaji and Ramnath Bhat.
\newblock Whatsapp vigilantes: An exploration of citizen reception and
  construction of whatsapp messages' triggering mob violence in india.
\newblock London School of Economics, 2019.

\bibitem[Pennycook and Rand(2019)]{pennycook2019lazy}
Gordon Pennycook and David~G Rand.
\newblock Lazy, not biased: Susceptibility to partisan fake news is better
  explained by lack of reasoning than by motivated reasoning.
\newblock \emph{Cognition}, 188:\penalty0 39--50, 2019.

\bibitem[Zauner(2010)]{zauner2010implementation}
Christoph Zauner.
\newblock Implementation and benchmarking of perceptual image hash functions.
\newblock 2010.

\bibitem[S{\"a}rndal et~al.(2003)S{\"a}rndal, Swensson, and
  Wretman]{sarndal2003model}
Carl-Erik S{\"a}rndal, Bengt Swensson, and Jan Wretman.
\newblock \emph{Model Assisted Survey Sampling}.
\newblock Science \& Business Media, 2003.

\bibitem[Lumley et~al.(2004)]{lumley2004analysis}
Thomas Lumley et~al.
\newblock Analysis of complex survey samples.
\newblock \emph{Journal of Statistical Software}, 9\penalty0 (1):\penalty0
  1--19, 2004.

\bibitem[Hemsley and Snyder(2018)]{hemsley2018dimensions}
Jeff Hemsley and Jaime Snyder.
\newblock Dimensions of visual misinformation in the emerging media landscape.
\newblock \emph{Misinformation and Mass Audiences (2018)}, 91, 2018.

\bibitem[Lee and Osindero(2016)]{lee2016recursive}
Chen-Yu Lee and Simon Osindero.
\newblock Recursive recurrent nets with attention modeling for ocr in the wild.
\newblock In \emph{CVPR}, 2016.

\bibitem[Chawla et~al.(2002)Chawla, Bowyer, Hall, and
  Kegelmeyer]{chawla2002smote}
Nitesh~V Chawla, Kevin~W Bowyer, Lawrence~O Hall, and W~Philip Kegelmeyer.
\newblock Smote: synthetic minority over-sampling technique.
\newblock \emph{Journal of AI research}, 2002.

\bibitem[Gupta et~al.(2013)Gupta, Lamba, Kumaraguru, and
  Joshi]{gupta2013faking}
Aditi Gupta, Hemank Lamba, Ponnurangam Kumaraguru, and Anupam Joshi.
\newblock Faking sandy: characterizing and identifying fake images on twitter
  during hurricane sandy.
\newblock In \emph{WWW}, pages 729--736. ACM, 2013.

\bibitem[Monti et~al.(2019)Monti, Frasca, Eynard, Mannion, and
  Bronstein]{monti2019fake}
Federico Monti, Fabrizio Frasca, Davide Eynard, Damon Mannion, and Michael~M
  Bronstein.
\newblock Fake news detection on social media using geometric deep learning.
\newblock \emph{arXiv preprint arXiv:1902.06673}, 2019.

\bibitem[Khattar et~al.(2019)Khattar, Goud, Gupta, and Varma]{khattar2019mvae}
Dhruv Khattar, Jaipal~Singh Goud, Manish Gupta, and Vasudeva Varma.
\newblock Mvae: Multimodal variational autoencoder for fake news detection.
\newblock In \emph{WWW}, pages 2915--2921. ACM, 2019.

\bibitem[Lokniti(2018)]{lokniti2018}
CSDS Lokniti.
\newblock How widespread is whatsapp's usage in india?, 2018.
\newblock URL
  \url{https://www.livemint.com/Technology/O6DLmIibCCV5luEG9XuJWL/How-widespread-is-WhatsApps-usage-in-India.html}.

\bibitem[Newman et~al.(2019)Newman, Fletcher, Kalogeropoulos, and
  Nielsen]{reuters2019report}
Nic Newman, Richard Fletcher, Antonis Kalogeropoulos, and Rasmus~Kleis Nielsen.
\newblock {Reuters Institute Digital News Report 2019 }.
\newblock Reuters Institute for the Study of Journalism, 2019.

\bibitem[Garimella and Tyson(2018)]{garimella2018whatsapp}
Kiran Garimella and Gareth Tyson.
\newblock Whatsapp, doc? a first look at whatsapp public group data.
\newblock In \emph{ICWSM}, pages 511--518, 2018.

\bibitem[Ester et~al.(1996)Ester, Kriegel, Sander, Xu,
  et~al.]{ester1996density}
Martin Ester, Hans-Peter Kriegel, J{\"o}rg Sander, Xiaowei Xu, et~al.
\newblock A density-based algorithm for discovering clusters in large spatial
  databases with noise.
\newblock In \emph{KDD}, pages 226--231, 1996.

\bibitem[Zannettou et~al.(2018)Zannettou, Caulfield, Blackburn, De~Cristofaro,
  Sirivianos, Stringhini, and Suarez-Tangil]{zannettou2018origins}
Savvas Zannettou, Tristan Caulfield, Jeremy Blackburn, Emiliano De~Cristofaro,
  Michael Sirivianos, Gianluca Stringhini, and Guillermo Suarez-Tangil.
\newblock On the origins of memes by means of fringe web communities.
\newblock In \emph{IMC}. ACM, 2018.

\end{thebibliography}
\end{document}